\documentstyle {report}
\begin{document}
\hfuzz=50pt
\newcommand{\be}{\begin{equation}}
\newcommand{\ee}{\end{equation}}
\newcommand{\bea}{\begin{eqnarray}}
\newcommand{\eea}{\end{eqnarray}}
\begin{titlepage}
\begin{center}
 {\Large \bf  Classical Gravity Coupled to Liouville Theory}\\
 \vspace{2cm}
 {\large \bf Francisco D. Mazzitelli\footnote{mazzitef@itsictp.bitnet}\\
\vspace{.5cm}
Noureddine Mohammedi}\footnote{Present address: University of Bonn,
D-5300, Bonn2}\\
\vspace{.5cm}
\large International Centre for Theoretical Physics \\
34100 Trieste, Italy.\\

\baselineskip 18pt

\vspace{.2in}

July, 1992 (revised version)\\
\vspace{1cm}
Abstract
\end{center}
We consider the two dimensional Jackiw-Teitelboim model of gravity.
We first couple the model to the Liouville action and
$c$ scalar fields and show, treating the
combined system as a non linear sigma model, that
the resulting theory can be interpreted as a critical string
moving in a target space of dimension $D=c+2$. We then analyse
perturbatively a generalised model containing a kinetic term
and an arbitrary potential for the auxiliary field.
We use the background field method and work with covariant
gauges. We show that the renormalisability of the theory depends
on the form of the potential. For a general potential, the theory
can be renormalised as a non linear sigma model.
In the particular case of a Liouville-like potential,
the theory is renormalisable in the usual sense.  \\

\end{titlepage}
\baselineskip 20pt
\parskip 16pt
\pagestyle{plain}
\pagenumbering{arabic}
\setcounter{chapter}{1}
\setcounter{section}{1}
\setcounter{section}{1}
\subsection*{1. Introduction}
\setcounter{equation}{0}
\setcounter{footnote}{0}
\par
There are two  motivations  behind  the recent intense activity in two
dimensional gravity. The first stream of thoughts sees in two dimensional
gravity a toy model for tackling the more subtle problem of four dimensional
gravity. The second considers two dimensional gravity as fundamental to the
study of string theory where one has to sum over all two dimensional
geometries. During the revival of string theory, however, the complication
of summing over geometries was spared by a restriction to certain critical
space-time dimensions and the Liouville mode was neglected [1].
Eventually, this
mode was included and new features appeared in the quantisation of string
theory [2-8]. Undoubtedly, the most stricking one is the fact that two
dimensional surfaces cannot be "embedded" in target spaces of dimensions
between 1 and 25. This is rather unnatural if the two dimensional surfaces
are to be thought of as world-sheets swept by the propagation and
interactions of strings in space-time.
\par
A major issue in the treatment of two dimensional gravity is in
finding a locally covariant action at the classical level. In the
pioneering work of Polyakov, this was a non-local functional which
reproduces the
well-know trace anomaly of the energy momentum tensor. How one analyses
this action in a general gauge is yet still unknown. The natural
analogue of the Einstein-Hilbert action is a topological invariant
counting the number of handles of the manifold. Nevertheless, some sense
can be made out of this action using perturbation theory and dimensional
regularisation [9], at least up to leading order [10]. An other
alternative for two dimensional
gravity was proposed by Jackiw and Teitelboim [11]. This expresses
the constancy
of the scalar curvature through the introduction of an extra field
(the dilaton) which seems
to spoil its geometric interpretation. It turns out that it is
this same field which makes
the quantisation of this theory much more interesting. Indeed,
it was shown
in ref.[12] that in this model the restriction on the
dimension of the target
space is completely lifted.
\par
In the present paper we analyse the Jackiw-Teitelboim model of two
dimensional gravity. Partial results of these analyses were reported
earlier in a short letter [13].  When coupled to $c$ scalar fields,
this model
behaves like a critical string in the sense that it forces $c$ to be
equal to 24. We then add a Liouville term to this model and treat
the resulting
theory as a  non-linear sigma model. The vanishing of the different beta
functions, up to linear terms in the tachyon field,
leads to the same results
obtained in ref.[12],  where the same model was considered as a theory of
non-critical strings. Therefore, this model can be interpreted also as a
theory of critical bosonic strings moving in a target space of dimension
$D=c+2$, where the Liouville mode and the extra field are string
coordinates too.

Then we consider a more general model which contains a kinetic term
and an arbitrary potential for the dilaton. We analyse this
model perturbatively and discuss its renormalisability. We show that
the theory is renormalisable in the usual sense for Liouville
(exponential) potentials. For other potentials, the theory can be
renormalised only in the sense of the non linear sigma models, that
is, allowing for  a change in the functional form of the potential.

The paper is organised as follows: In section two we study the
Jackiw-Teitelboim model together with the Liouvillle action using
conformal field theory techniques. We then add matter fields and
treat the whole theory as a non-linear sigma model.
In the third section we consider the above mentioned generalised
model. We expand
the action up two second order in the quantum fields of the
background field expansion and choose our gauge fixing terms and
calculate their corresponding ghost action. In the fourth section
we present our results for the one loop divergences using a generalised
formula of the heat-kernel method. The proof of this formula is given
in an appendix. Section five deals with the renormalisation
procedure. Finally, we end our
article with some concluding remarks.

\setcounter{chapter}{2}
\setcounter{section}{1}
\setcounter{section}{1}
\subsection*{2. The Jackiw-Teitelboim model coupled to
Liouville }
\setcounter{equation}{0}

The classical gravity action is assumed to be given by
\be
S_{JT}={b\over \pi}\int d^2 x\sqrt g N (R+\Lambda)\,\,\,,
\ee
where $b$ is a constant and $N(x)$ is an auxiliary field whose equation
of motion yields the Einstein-like equation in two dimensions
\be
R+\Lambda=0\,\,\,.
\ee
This action was first proposed by Jackiw and Teitelboim [11]
as an alternative
to the usual Einstein-Hilbert action,  $\int d^2x\sqrt g R$,
which is trivial
in two dimensions. Many interesting aspects of this
model were considered
in refs.[14-18].

In the background geometry specified by $\hat g_{\alpha\beta}$, where
\be
g_{\alpha\beta}=\hat g_{\alpha\beta} e^{\gamma\sigma}\,\,\,,\,\,\,
\gamma=1\,\,or\,\,2\,\,\,,
\ee
the above action becomes
\be
S_{JT}={b\over \pi}\int d^2 x\sqrt{\hat g}
[\gamma\hat g^{\mu\nu}\partial_{\mu}
\sigma\partial_{\nu} N + N(\hat R+\Lambda e^{\gamma\sigma})]\,\,\,,
\ee
where we have used the well-known result
\bea
R&=&e^{-\gamma\sigma}(\hat R-\gamma\nabla^2_{\hat g}\sigma)\nonumber\\
\nabla^2_{\hat g}&=&{1\over\sqrt{\hat g}}\partial_{\alpha}(\sqrt{\hat g}
\hat g^{\alpha\beta}\partial_{\beta})\,\,\, .
\eea
The energy momentum tensor corresponding to $S_{JT}$ with $\Lambda=0$
is given by
\be
T_{\alpha\beta}=-{4\pi\over\sqrt{\hat g}}{\delta L\over\delta\hat g^{\alpha
\beta}}=-4b[\gamma\partial_{(\alpha}\sigma\partial_{\beta )}N -
\hat\nabla_{\alpha}
\hat\nabla_{\beta}N + \hat g_{\alpha\beta}(-{\gamma\over 2}\hat
g^{\mu\nu}\partial_{\mu}
\sigma\partial_{\nu}N+\nabla ^2 N)]\,\,\,.
\ee
The z-z component of this energy momentum tensor is written as
\be
T_{zz}=-4b(\gamma\partial_z\sigma\partial_z N-\partial^2_zN)\,\,\,.
\ee
The only propagator of this theory is
\be
<\sigma (z) N(w)>=-{1\over 4\gamma b} ln(z-w)\,\,\,.
\ee
The operator product expansion of the energy momentum tensor produces a
central charge for gravity given by
\be
c^{gravity}=2\,\,\,.
\ee
If we introduce matter interactions through an action for $c$ scalar fields
$X^i$
\be
S_{matter}={1\over 4\pi}\int d^2x\sqrt {\hat g}\hat g^{\mu\nu}\partial_
{\mu}X^i\partial_{\nu}X^i\,\,,\,\,i=2,\dots ,c+1
\ee
then their contribution to the total central charge, together with that
of the ghosts, is $c-26$. Hence requiring that the total central charge
vanishes leads to $c=24$ ! The obvious question to be asked now is
whether it is possible to couple matter fields to our theory when
$c\neq 24$.

The action $S_{JT}$ can be modified in two ways: The first one consists
in adding a kinetic term for the field $N$. This, however, results in a
change in equation (2.2) which is the main motivation for proposing (2.1) as
an action for classical gravity in two dimensions. The other alternative
which will be adopted in this paper is to add a Liouville
action
\be
S_L={1\over \pi}\int d^2 x\sqrt{\hat g}(a\hat g^{\mu\nu}\partial_{\mu}
\sigma\partial_{\nu}\sigma  + Q\sigma {\hat R})
\ee
to our action in (2.1), where $a$ and $Q$ are two constants. The resulting
action $S_{tot}=S_{JT}+S_L+S_{matter}$,
is  proportional to the action for two dimensional gravity interacting with
matter
fields proposed in ref.[12].
Standard analyses show that the energy momentum tensor corresponding
to $S_{JT}+S_L$ has a central charge given by
\be
c^{gravity}=2+96\left({Q\over \gamma}-{a\over\gamma^2}\right)\,\,\,.
\ee
Including the matter and ghost contributions, the vanishing of
the total central charge leads to
\be
c-96\left({1\over 4}-{Q\over\gamma}+{a\over\gamma^2}\right)=0\,\,\,.
\ee
Unlike the Liouville alone [2-4], there is no restiction on the matter
central charge.

Notice that $S_{tot}$ can be written as a non-linear sigma model in some
special backgrounds. This is given by
\be
S_{tot}={1\over 4\pi}\int d^2x\sqrt{\hat g}[G_{ab}(\eta)\hat g^{\mu\nu}
\partial_{\mu}\eta^a\partial_{\nu}\eta^b + \hat R\Phi(\eta)
+2\mu T(\eta)]\,\,\,.
\ee
Here $\eta^0=\sigma$,$\eta^1=N$ and $\eta^{a}=X^i$ for $ a=2,...,c+1$.
The target space metric is given by $G_{00}=4a, G_{01}=2\gamma b, G_{11}=0,
G_{ab}=\delta_{ij}$ for $a,b=2,...c+1$ and $G_{0i}=G_{1i}=0$.
The dilaton field is
\be
\Phi(\eta)=4Q\sigma+4bN\,\,\,.
\ee
We have also included
a cosmological term in the form of a tachyon
\be
T(\eta)=e^{\alpha\sigma}\,\,\,,
\ee
where $\alpha$ is a constant to be determined.

To linear terms in the tachyon $T$, the vanishing of the $\bar\beta$
functions leads to \footnote {Our conventions for the
$\beta$-functions are those of ref.[19], and we have included
only the terms that are relevant to our calculation.}
\bea
\bar\beta^G_{ab}&=&R_{ab}+2\nabla_a\nabla_b\Phi+...=0\\
\bar\beta^{\Phi}&=&{1\over 6}(D-26)-{1\over 2}\nabla^2\Phi+G^{ab}\nabla_a\Phi
\nabla_b\Phi+...=0\\
\bar\beta^T&=&-{1\over 2}G^{ab}\nabla_a\nabla_bT+G^{ab}\nabla_a\Phi
\nabla_bT-2T+...=0
\eea
In $\bar\beta^{\Phi}$, $D=c+2$ is the dimension of the target space (or
the number of fields). With
the above backgrounds $\bar\beta^G_{ab}=0$ is automatically satisfied,
whereas
the equations for $\bar\beta^{\Phi}$ and $\bar\beta^T$ lead respectively
to
\bea
&{1\over 6}(c-24)+16({Q\over\gamma}-{a\over\gamma^2})=0&\\
&\alpha =\gamma&
\eea
The last equation is just the requirement that $T(\eta)$ is a (1,1)
operator with respect to the energy momentum tensor corresponding
to $S_{JT}+S_L$ [12]. Notice that Eq. (2.10) is exactly equivalent to
Eq. (2.13) obtained by conformal field theory considerations. Therefore
$S_{tot}$, which is a theory of
non-critical strings, can be interpreted as a theory of a critical bosonic
string
moving in a target space of dimension $D=c+2$.
In particular, for $Q=a$ and $\gamma =2$ we have
\be
a={1\over 24}(24-c)
\ee

This is also the result obtained in ref.[12] utilising conformal field theory
methods.
\footnote{After completion of this work we learned that the sigma-model
interpretation
of the action $S_{tot}$ has been treated by Chamseddine in Ref. [20]}

\setcounter{chapter}{3}
\setcounter{section}{1}
\setcounter{section}{1}

\subsection*{3.Perturbation theory}

\setcounter{equation}{0}

We will use the background field expansion and dimensional
regularisation
where $d=2-\epsilon$.
In this section we will find the expansion up to second order
in the quantum fields
of the classical
action. Our starting point is
\be
S=\int d^d x\sqrt{\bar g}\left( \bar N \bar R +V(\bar N)+ {G\over 2}
\partial _{\mu}\bar N\partial ^{\mu}\bar N +{1\over 2}
\partial _{\mu}
\bar X^i\partial ^{\mu}\bar X^i\right )\,\,\,,
\ee
where we  have  included
a kinetic term and an arbitrary  potential for the dilaton field.
Here G is an arbitrary  constant.
For $V(\bar N)=0$ the first two terms of the
action are the local counterpart of the non-local Polyakov action
$S=\int d^dx\sqrt g R(\nabla^{2}) ^{-1} R$. \footnote{This
non-local action has been also studied perturbatively
in ref.[21]} Similar models have
been considered in ref. [22].

The classical equations of motion are given by
\bea
&R+V'-G\nabla^2 N=0&\\
&\nabla^2N g_{\mu\nu}-N_{;\mu\nu}+N[R_{\mu\nu}-
{1\over 2}Rg_{\mu\nu}]-{1\over 2}
Vg_{\mu\nu}=&\nonumber\\
&{G\over 2}[{1\over 2}g_{\mu\nu}N_{;\rho}N^{;\rho}-
N_{;\mu}N_{;\nu}]
+{1\over 2}[{1\over 2}g_{\mu\nu}X_{;\rho}^iX^{i;\rho}-X^i_{;\mu}
X^i_{;\nu}]&\\
&\nabla^2 X^i=0\,\,\, .&
\eea
Note that eq. (3.3) implies that $\nabla^2N=V+O(\epsilon)$.

To compute the one loop effective action we use the background field
method, expanding the full fields $\bar g_{\mu\nu},\bar N$ and
$\bar X^i$ around the classical configurations $g_{\mu\nu}, N$ and
$X^i$, that is
\bea
\bar g_{\mu\nu}&=&g_{\mu\nu}+h_{\mu\nu}\nonumber\\
\bar N&=& N+\varphi\nonumber\\
\bar X^i&=&X^i+\xi^i\,\,\,,
\eea
where $h_{\mu\nu}, \varphi$ and $\xi^i$ are quantum fluctuations.
Dropping the linear terms proportional to the classical equations of
motion we get
\bea
&S&(\bar g_{\mu\nu},\bar N, \bar X^i)-S(g_{\mu\nu},N,X^i)=S^{(2)}=
\nonumber\\
&=&{1\over 2}\int d^dx\sqrt g N ({1\over 2}h_{;\mu}h^{;\mu}
-h^{;\mu}h_{\mu\nu}^{;\nu}-{1\over 2}h_{\mu\nu ;\rho}
h^{\mu\nu;\rho}
+h_{\mu\nu ;\rho}h^{\mu\rho ;\nu})\nonumber\\
&+&{1\over 2}\int d^dx\sqrt g N(-{1\over 2}Rh_{\mu\nu}h^{\mu\nu}
+{1\over 4}Rh^2-hh^{\mu\nu}R_{\mu\nu}+2h^{\mu\rho}h_{\rho}^{\nu}
R_{\mu\nu})
\nonumber\\
&+&\int d^dx\sqrt g\varphi ( {1\over 2}Rh - R_{\mu\nu}h^{\mu\nu}
-\nabla^2h + h^{\mu\nu}_{\,\,\, ;\mu\nu} )\nonumber\\
&+&\int d^dx\sqrt g[{1\over 2}\nabla^2N(h_{\mu\nu}h^{\mu\nu}-{1\over 2}
h^2)+N_{;\mu\nu}(hh^{\mu\nu}-h^{\mu}_{\rho}h^{\rho\nu})
+{1\over 2}N_{;\mu}hh^{\mu\nu}_{\,\,\, ;\nu}]\nonumber\\
&+&\int d^dx\sqrt g[{V\over 4} ( {1\over 2}h^2 -
h_{\mu\nu}h^{\mu\nu} ) +{1\over 2}V'\varphi h +{1\over 2}
V''\varphi ^2]\nonumber\\
&+&{G\over 2}\int d^dx\sqrt g [\varphi_{;\mu}\varphi^{;\mu} + N_{;\mu}
N_{;\nu} ( -{1\over 2}hh^{\mu\nu}+h^{\mu\rho}h_{\rho}^{\nu}\nonumber\\
&+&{1\over 4}g^{\mu\nu} ({h^2\over 2}-h_{\rho\sigma}h^{\rho\sigma}
 )  ) + N_{;\mu}\varphi_{;\nu} (hg^{\mu\nu}-2h^{\mu\nu} )
 ]\nonumber\\
&+&{1\over 2}\int d^dx\sqrt g [\xi_{;\mu}^i\xi^{i;\mu} +
X_{;\mu}^i\xi_{;\nu}^i  ( hg^{\mu\nu} - 2h^{\mu\nu} )\nonumber\\
&+& X^i_{;\mu}X^i_{;\nu} ({1\over 4}g^{\mu\nu} ({1\over 2} h^2 -
h_{\rho\sigma}h^{\rho\sigma}) - {1\over 2}hh^{\mu\nu}+ h^{\mu\rho}h_{\rho}
^{\nu})]\,\,\,,
\eea
where $h\equiv h^\mu_{\,\,\,\mu}$.

To proceed, we must add a gauge fixing Lagrangian. We will choose a gauge
in such a way that
the differential operator in the kinetic term is always
$\nabla^2=\nabla_{\mu}\nabla^{\mu}$. Having a  kinetic
term proportional to the Laplacian, the evaluation of the one loop
divergences becomes simpler, as one has to compute the determinant of
a minimal operator. Standard heat-kernel techniques are then applicable [23].
In usual gravity the DeWitt gauge does the job [24]. A suitable generalisation
of this gauge to our model is found to be
\be
S_{gf}=-{1\over 2}\int d^dx\sqrt g N[h_{\nu ;\mu}^{\mu}-{1\over 2}
h_{;\nu}-{1\over N}\varphi_{;\nu}-\beta (N)N^{;\mu}h_{\mu\nu}]^2\,\,\,\, ,
\ee
where $\beta (N)$ is an arbitrary function
of the classical field $N$.

The corresponding ghost action for this gauge condition is
\bea
S_{gh}&=&\int d^dx\sqrt g\bar C^{\mu}\Delta_{\mu\nu}^{gh} C^{\nu}\nonumber\\
\Delta_{\mu\nu}^{gh}&=&N[g_{\mu\nu}\nabla^2-{1\over N}N_{;\nu}\nabla_{\mu}
-\beta(N_{;\nu}\nabla_{\mu}+g_{\mu\nu}N^{;\rho}\nabla_{\rho})\nonumber\\
&-&{1\over N}N_{;\nu\mu}+R_{\mu\nu}]
\,\,\, .
\eea

Adding the gauge fixing term (3.7) to the quadratic Lagrangian (3.6)
we get, in a condensed notation,
\bea
S_T^{(2)}=S^{(2)}+S_{gf}&=&{1\over 2}\int d^dx\sqrt g h_{mn}(-\Delta^{mn,pq}
\nabla^2+Y_{\mu}^{mn,pq}\nabla^{\mu}
\nonumber\\&+&X^{mn,pq}-{1\over 2}\nabla^{\mu}
S_{\mu}^{mn,pq})h_{pq}\,\,\, .
\eea
Here the indices $mn,pq$ run from 1 to d+1+c. The field $h_{mn}$ coincides
with $h_{\mu\nu}$ for $m,n=1,...d$, while $h_{d+1\,\,d+1}\equiv
h_{\varphi\varphi}$
is defined to be equal to $\varphi$ and $h_{d+1+i\,\,\,d+1+i}\equiv h_{ii}$
denotes the
matter field $X^i$. The pairs $(mn)$ and $(pq)$ take the values $(\mu\nu)$,
$(d+1\,\,d+1)\equiv (\varphi\varphi)$ or $(d+1+i\,\,d+1+i)\equiv (ii)$ but
crossed pairs like
$(\mu\varphi)$,$(\mu i)$ and $(i\varphi)$ must not be included. The matrices
$\Delta$, $Y_{\mu}$, $S_{\mu}$ and $X$ are given by
\be
\Delta^{mn,pq}=\left\{\begin{array}{ll}
\Delta^{ii,ii}=1\\
\Delta^{\varphi\varphi ,\varphi\varphi}=G-{1\over N}\\
\Delta^{\varphi\varphi ,\mu\nu}=\Delta^{\mu\nu ,\varphi\varphi}={1\over
2}g^{\mu\nu}
\\
\Delta^{\mu\nu ,\rho\sigma}=N P^{\mu\nu ,\rho\sigma}\\
\end{array}
\right.
\ee

\be
X^{mn,pq}=\left\{\begin{array}{ll}
X^{ii,ii}=X^{\varphi\varphi ,\varphi\varphi}=0\\
X^{\mu\nu ,\varphi\varphi}=X^{\varphi\varphi ,\mu\nu}=
-R^{\mu\nu}+{1\over 2}Rg^{\mu\nu}+{V'\over 2}g^{\mu\nu}\\
X^{\mu\nu ,\rho\sigma}=N T^{\mu\nu ,\rho\sigma}+G[{1\over 2 }
N_{;\tau}
N^{;\tau} P^{\mu\nu ,\rho\sigma}-{1\over 4}(g^{\mu\nu}N^{;\rho}
N^{;\sigma}
+g^{\rho\sigma}N^{;\mu}N^{;\nu})]\\
\,\,\,\,\,\,+{1\over 2} (G-\beta^2 N)[N^{;(\mu}N^{;\sigma}g^{\nu )\rho }+
N^{;(\mu}N^{;\rho}g^{\nu )\sigma }]\\
\,\,\,\,\,\,
-2P^{\mu\nu ,\rho\sigma}\nabla^2N+{1\over 4}(N^{;\mu\nu}g^{;\rho\sigma}
+N^{;\rho\sigma}g^{\mu\nu})\\
\,\,\,\,\,\,
+{1\over 2}\partial_{\alpha}X^i\partial^{\alpha}X^i
P^{\mu\nu ,\rho\sigma}
-{1\over 4}(g^{\mu\nu}\partial^{\rho}X^i\partial^{\sigma}X^i
+g^{\rho\sigma}\partial^{\mu}X^i\partial^{\nu}X^i)\\ \,\,\,\,\,\,
+{1\over 2}[ g^{\rho (\mu}\partial^{\sigma}X^i\partial^{\nu
)}X^i+g^{\sigma(\mu}\partial^{\rho}X^i\partial^{\nu )}X^i]
\end{array}
\right.
\ee

\be
S_{\tau}^{\, mn,pq}=\left\{\begin{array}{ll}
S_{\tau}^{\,\varphi\varphi ,\varphi\varphi}=
-{1\over N^2}N_{;\tau}\\
S_{\tau}^{\,\varphi\varphi ,\rho\sigma}=S_{\tau}^{\,\rho\sigma ,
\varphi
\varphi}=2GN_{;\nu}P^{\rho\sigma ,\nu}_{\,\,\,\,\,\,\,\,\tau}
-{1\over 2}\beta
(N^{;\rho}g^{\sigma}_{\tau}+N^{;\sigma}g^{\rho}_{\tau})\\
S_{\tau}^{\,\mu\nu ,\rho\sigma}=-N_{;\tau}P^{\mu\nu ,\rho\sigma}
-(N\beta+2)
(N^{;\mu}P^{\rho\sigma ,\nu}_{\,\,\,\,\,\,\,\,\tau}\\
\,\,\,\,\,\,\,\,\,\,\,\,\,\, + N^{;\nu}P^{\rho\sigma
,\mu}_{\,\,\,\,\,\,\,\,\,\tau}+N^{;\rho}P^{\mu\nu ,\sigma}
_{\,\,\,\,\,\,\,\,\tau}+N^{;\sigma}
P^{\mu\nu ,\rho}_{\,\,\,\,\,\,\,\,\tau})\\
S_{\tau}^{\,\rho\sigma  ,ii}=S_{\tau}^{\, ii
,\rho\sigma}=2X^i_{;\nu}P^{\rho\sigma ,\nu}_{\,\,\,\,\,\,\,\,\tau}\\
\end{array}
\right.
\ee

\be
Y_{\tau}^{mn,pq}=\left\{\begin{array}{ll}
Y_{\tau}^{\varphi\varphi ,\varphi\varphi}=0\\
Y_{\tau}^{\,\,\rho\sigma ,\varphi\varphi}=-
Y_{\tau}^{\,\,\varphi\varphi ,
\rho\sigma}=S_{\tau}^{\,\,\rho\sigma ,\varphi\varphi}\\
Y_{\tau}^{\,\,\rho\sigma ,ii}=-Y_{\tau}^{\,\, ii,\rho\sigma}=
S_{\tau}^{\,\,
\rho\sigma ,ii}\\
Y_{\tau}^{\,\,\mu\nu ,\rho\sigma}=(\beta N+1)
 (N^{;\rho}P^{\mu\nu ,\sigma}_
{\,\,\,\,\,\,\,\,\tau}+N^{;\sigma}P^{\mu\nu
,\rho}_{\,\,\,\,\,\,\,\,\tau}-N^{;\mu}
P^{\rho\sigma ,\nu}_{\,\,\,\,\,\,\,\,\tau}-
N^{;\nu}P^{\rho\sigma ,\mu}_{\,\,\,\,\,\,\,\,\tau})\\
\end{array}
\right.
\ee
where
\bea
P^{\mu\nu ,\rho\sigma}&=&{1\over 4}(g^{\mu\nu}g^{\rho\sigma}
-g^{\mu\rho}g^{\nu\sigma}-g^{\mu\sigma}g^{\nu\rho})\\
I^{\mu\nu ,\rho\sigma}&=&{1\over 2}(g^{\mu\rho}g^{\nu\sigma}+
g^{\mu\sigma}g^{\nu\rho})
\eea
and
\bea
T^{\mu\nu ,\rho\sigma}&=&{1\over 4}(R+{V\over N} )
(g^{\mu\nu}g^{\rho\sigma}-g^{\mu\rho}g^{\nu\sigma}-g^{\mu\sigma}g^{\nu\rho})
\nonumber\\
&+&{1\over 4}(g^{\mu\rho}R^{\nu\sigma}+g^{\mu\sigma}R^{\nu\rho}
+g^{\nu\rho}R^{\mu\sigma}+g^{\nu\sigma}R^{\mu\rho})\nonumber\\
&-&{1\over 2}(g^{\mu\nu}R^{\rho\sigma}+g^{\rho\sigma}R^{\mu\nu})
+{1\over 2}(R^{\nu\rho\mu\sigma}+R^{\nu\sigma\mu\rho})\,\,\,,
\eea
The components of $\Delta^{-1}$ are given by
\be
(\Delta^{-1})_{pq,rs}=\left\{\begin{array}{ll}
\Delta^{-1}_{ii,ii}=1\\
\Delta^{-1}_{\varphi\varphi ,\varphi\varphi}=0\\
\Delta^{-1}_{\varphi\varphi ,\mu\nu}=\Delta^{-1}_{\mu\nu ,
\varphi\varphi}=
g_{\mu\nu}\\
\Delta^{-1}_{\mu\nu ,\rho\sigma}={2\over N}[-I_{\mu\nu ,\rho\sigma}
+g_{\mu\nu}g_{\rho\sigma}(1-{1\over 2}NG)]\,\,\, .\\
\end{array}
\right.
\ee
Note that, unlike the pure gravity case [9], the
inverse matrix $\Delta^{-1}$
is well defined and has no pole in the limit
$\epsilon\rightarrow 0$. This is
of course due to the fact that the action
(3.1) is non trivial in exactly
two dimensions.
This fact is also crucial for the absence of
any renormalisation
for the anomalous dimensions of operators [9].
We also stress that the structure of the propagator here
is different from that of the conformal gauge analysis
in section 2. There, after adding the Liouville action,
one has $<NN>\neq 0, <N\sigma >\neq 0$ and $<\sigma\sigma>=0$
(see also ref. [20]). This result does not appear in
Eq. (3.17) in the limit $G=0$. The reason for this apparent
discrepancy is that the gauge fixing Lagrangian induces
new kinetic terms for the field $\varphi$ and thus
modifies the structure of the propagator of the theory.

In deriving Eqs.(3.9)-(3.16) we performed some integrations
by parts in such a way that
the matrices $\Delta$,$X$ and $S_{\mu}$ are symmetric and $Y_{\mu}$
is antisymmetric under the interchange of the pairs $(mn)$ and $(pq)$.
This property will be important when computing the divergences.

\setcounter{chapter}{4}
\setcounter{section}{1}
\setcounter{section}{1}

\subsection*{4.The one loop divergences}

\setcounter{equation}{0}

To compute the one loop effective action, we need to evaluate the functional
determinant of the (symmetrized) differential operators that appear in
Eqs.(3.8) and (3.9). Using heat-kernel techniques [23], it can be shown
that given the operator
\be
[-\nabla^2 I_{ij}+A^{\mu}_{ij}\nabla_{\mu} + M_{ij}]
\ee
the logarithm of its determinant is given by
\bea
&ln\,\, det
[-\nabla^2 I_{ij}+A^{\mu}_{ij}\nabla_{\mu} + M_{ij}]
= {1\over 2\pi\epsilon}Tr\left ( -{1\over 6}R I+ {1\over 4} A_{\mu}A^{\mu}
+M\right ) &\nonumber\\
&= {1\over 2\pi\epsilon}\int d^dx\sqrt g\left (-{1\over 6}R  I_{ii} +
{1\over 4}A_{\mu\, ij}A^{\mu}_{ji} + M_{ii}\right ) \,\,\,,&
\eea
where $\epsilon=d-2$ and $I_{ij}$ is the identity operator.
The proof of this formula is presented in the
Appendix A.

In our case, there is an additional complication because the Laplacian
in the kinetic term is multiplied by a space-time dependent matrix
which does not commute with the covariant derivative. To get rid of
this we use the doubling trick of t'Hooft and Veltman [25]. We consider
the action (3.9) and add the same expression but with different fields
$h'_{mn}$ instead of $h_{mn}$. The `doubled' effective action will be
two times the original effective action. In terms of the complex fields
\bea
\lambda_{mn}=h_{mn}+ih'_{mn}\nonumber\\
\bar\lambda_{mn}=h_{mn} - ih'_{mn}
\eea
we can write the doubled action as
\bea
S_T^{(2)}&=&{1\over 2}\int d^dx\sqrt g\bar\lambda_{mn} (-\Delta^{mn,pq}
\nabla^2 +(Y_{\mu}^{mn,pq}-\nabla_{\mu}\Delta^{mn,pq})\nabla^{\mu}
\nonumber\\
&+&X^{mn,pq}-{1\over 2}\nabla^{\mu}S_{\mu}^{mn,pq}+{1\over 2}\nabla^{\mu}
Y_{\mu}^{mn,pq}-{1\over 2}\nabla^2\Delta^{mn,pq})\lambda_{pq}
\eea
where we have used the symmetry and antisymmetry properties of $\Delta, X,
S_{\mu}$ and $Y_{\mu}$ respectively. As we are now dealing with
complex fields,
the one loop divergences remain unchanged under the replacement [25]
\be
\bar\lambda_{rs}\rightarrow\Delta^{-1}_{rs,mn}\bar\lambda^{mn}\,\,\,.
\ee
As a consequence, the effective action associated with the action (3.9) is
\bea
W^{(1)}&=&-{1\over 2} ln\,\,det [-\nabla^2 I_{rs}^{\,\,\,\,\,\, pq} +
\Delta^{-1}_{rs,mn}
(Y_{\mu}^{mn,pq}-\nabla_{\mu}\Delta^{mn,pq})\nabla^{\mu}\nonumber\\
&+&\Delta^{-1}_{rs,mn}(X^{mn,pq}-{1\over 2}\nabla ^{\mu}S_{\mu}^{mn,pq}
+{1\over 2}\nabla^{\mu}Y_{\mu}^{mn,pq}-{1\over 2}\nabla^2\Delta^{mn,pq})]
\eea
where we have included a factor ${1\over 2}$ to take into account
the doubling
of the degrees of freedom.

A replacement analogous to Eq.(4.5) can be done for the ghost fields in
order to get rid of the $Ng_{\mu\nu}$ factor appearing in the kinetic term.
We have then
\be
W^{(1)}_{gh}=ln\,\,det N^{-1}g^{\nu\rho}\Delta^{gh}_{\mu\nu}\,\,\,,
\ee
where now the coefficient in front of $ln \,\,det$ is $+1$ due to the
fact that the ghosts are complex anticommuting fields.

Now we are ready to compute the total one loop effective action
\be
W^{(1)}_T=W^{(1)}+W^{(1)}_{gh}\,\,\,.
\ee
All we have to do is to apply the heat-kernel formula (4.2) to evaluate
the determinants. We begin with the ghosts contribution.
According to Eqs.(3.8), (4.2) and (4.7) we have
\bea
W^{(1)}_{gh}&=&ln\, \,\,det [g^{\rho}_{\mu}\nabla^2 -A^{\tau\,\,\rho}_{\,\,\mu}
\nabla_{\tau}+R^{\rho}_{\mu}-{1\over N}N_{;\mu}^{\,\,;\rho}]\nonumber\\
&=&{1\over 2\pi\epsilon}\int d^dx\sqrt g[-{4\over 3}R+{1\over
4}A^{\tau\,\,\rho}
_{\,\,\mu}A_{\tau\,\,\rho}^{\,\,\,\mu}+{1\over N}\nabla^2 N]
\eea
where
\be
A^{\tau\,\,\rho}_{\,\,\mu}={1\over N}N^{;\rho}g^{\tau}_{\mu}+
\beta (N^{;\rho}g^{\tau}_{\mu}+N^{;\tau}g^{\rho}_{\mu})
\ee
A simple calculation gives
\be
A^{\tau\,\,\rho}_{\,\,\mu}A_{\tau\,\,\rho}^{\,\,\mu}=
{1\over N^2}N_{;\rho}N^{;\rho}(1+4N\beta + 5N^2\beta ^2)
\ee
so, after integrations by parts we obtain
\be
W^{(1)}_{gh}={1\over 4\pi\epsilon}[-{8\over 3}R+{1\over 2}
{N_{;\rho}N^{;\rho}\over N^2}(5+4N\beta+5N^2\beta^2)]
\ee

Now we consider the calculation of $W^{(1)}$. From Eqs.(4.2) and (4.6)
we have
\bea
W^{(1)}&=&-{1\over 2}\ln\, det[-\nabla^2 I_{rs}^{\,\,pq}+\tilde
Y_{rs}^{\tau\,\,\,
pq}\nabla_{\tau}+\tilde X_{rs}^{\,\,\,pq}]\nonumber\\
&=&{1\over 4\pi\epsilon}\int d^dx\sqrt g [{(c+4)\over 6}R-{1\over 4}
\tilde Y^{\tau\,\,\, pq}_{rs}\tilde Y_{\tau\,\,\, pq}^{\,\,\,\,\,\,\,\, rs}
-\tilde X_{rs}^{\,\,\, rs}]
\eea
where
\bea
\tilde X_{rs}^{\,\,\,pq}&=&\Delta^{-1}_{rs,mn}(X^{mn,pq}
+{1\over 2}\nabla^{\mu}Y_{\mu}^{mn,pq}
-{1\over 2}
\nabla^{\mu}S_{\mu}^{mn,pq}-{1\over 2}\nabla^2\Delta^{mn,pq})\\
\tilde Y^{\tau\,\,\,\,pq}_{rs}&=&\Delta^{-1}_{rsmn}(Y_{\tau}^
{mn,pq}-\nabla_{\tau}\Delta^{mn,pq})
\eea
\par
The trace of the $X$-term is given by
\bea
\tilde X_{rs}^{\,\,\, rs}&=&2R+2V'+2{V\over N}+{N_{;\tau}N^{;\tau}\over
N^2}((\beta N+1)^2+GN(\beta^2 N^2-1))\nonumber\\
&-&{1\over N}X^{i;\tau}X_{;\tau}^i
\eea
\par
The next step is to compute the $\tilde Y^2$-trace in Eq.(4.13).
The following
identities are useful in this calculation
\bea
&P^{\mu\nu ,\rho}_{\,\,\,\,\,\,\,\,\,\rho}=0\,\,\,\, ,\,\,\,
P^{\mu\nu ,}_{\,\,\,\,\,\,\mu\nu}=-1
\,\,\, ,\,\,\, P^{\mu\nu ,\sigma}_{\,\,\,\,\,\,\,\nu}=
-{1\over 2}g^{\mu\sigma}&
\nonumber\\
&I_{\mu\nu ,\rho}^{\,\,\,\,\,\,\,\rho}=g_{\mu\nu}\,\,\,\,\, ,\,\,\,\,\,
I_{\mu\nu ,}^{\,\,\,\,\,\,\,
\nu\sigma}={3\over 2}g^{\sigma}_{\mu}&\nonumber\\
&P_{\mu\nu ,\tau}^{\,\,\,\,\,\,\,\,\,\epsilon}P^{\mu\nu
,\tau}_{\,\,\,\,\,\,\,\,\,\sigma}
={1\over 4}g^{\epsilon}_{\sigma}\,\,\,\, ,\,\,\,\,
I_{\rho\sigma}^{\,\,\,\mu\nu}
I^{\rho\sigma\,\,\tau}_{\,\,\,\,\mu}={3\over 2}g^{\nu\tau}&
\eea
We are not  including $O(\epsilon)$ terms because they produce
finite contributions to the effective action.
According to Eq.(4.15) we have that
\bea
tr\,\tilde
Y^2&=&\Delta^{-1}_{rs,mn}(Y_{\tau}^{mn,pq}-\nabla_{\tau}\Delta^{mn,pq})
\Delta^{-1}_{pq,tu}(Y_{\tau}^{tu,rs}-\nabla_{\tau}\Delta^{tu,rs})
\nonumber\\
&=&\Delta^{-1}_{rs,mn}\nabla_{\tau}\Delta^{mn,pq}\Delta^{-1}_{pq,tu}
\nabla_{\tau}\Delta^{tu,rs}+\Delta^{-1}_{rs,mn}Y_{\tau}^{mn,pq}
\Delta^{-1}_{pq,tu}Y_{\tau}^{tu,rs}
\eea
Note that due to the symmetry of $\Delta^{-1}$ and the
antisymmetry of $Y$ the crossed terms do not contribute.  Using
the cyclic property of the
trace an neglecting total derivatives we get
\be
tr\,\Delta^{-1}(\nabla_{\tau}\Delta)\Delta^{-1}(\nabla^{\tau}\Delta)=
2{N_{;\rho}N^{;\rho}\over N^2}
\ee
The result for the second term in (4.18) is
\bea
tr\,\Delta^{-1}Y\Delta^{-1}Y&=&{4\over N}X^i_{;\tau}
X^{i;\tau}+{N_{;\tau}N^{;\tau}\over N^2}\times\nonumber\\
\times [2\beta^2N^2&+&8N(\beta N+1)(\beta+G)
-4(\beta N +1)^2(1+GN)]
\eea

Combining the above equations we find, finally,
\be
W^{(1)}_T= {1\over 2\pi\epsilon}\int d^dx\sqrt g[{(c-24)\over 12}R
-V' -{V\over N} + {\nabla^2N\over N}]\,\,\,\,   ,
\ee
where we dropped a boundary term.

Several comments are in order. First of all, we see that the divergences
are not of the form of the classical action. As usual in quantum gravity,
a field redefinition [25,26] will  be necessary to
renormalise the theory.
However, as we will see in the next section, the theory is
renormalisable in the usual sense only for a particular class
of potentials. It is also worth noting that the arbitrary
function $\beta (N)$ introduced by the gauge fixing term has
disappeared from our final answer, as well as  the constant $G$
and the classical
matter fields $X^i$.
Other interesting feature is that the coefficient of $R$ in
the one loop divergence reproduces exactly the result for the
coefficient $a$ appearing in section 2 (see eq. (2.22)).

\setcounter{chapter}{5}
\setcounter{section}{1}
\setcounter{section}{1}

\subsection*{ 5.Renormalisation}

\setcounter{equation}{0}

The usual renormalisation procedure would be to absorb the infinities
into the bare constant $G$ and the bare constants appearing in
the potential $V(N)$,
allowing  for a field redefinition of $N$ and $g_{\mu\nu}$.
We will follow here a generalised procedure, allowing also for
a change of the functional form of the potential $V(N)$. This is
similar to what is done when renormalising a non-linear sigma model
in two spacetime dimensions.

In our divergence (4.21), the combination of the last two terms
vanishes on shell because it is the trace of the classical
equation (3.3). This means that these terms can be absorbed into
a conformal rescaling of the metric. The term proportional
to the curvature $R$ can be absorbed into a constant shift of
the scalar field $N$. As a consequence, one way of absorbing
the infinities is
\bea
N&\rightarrow & N-{1\over 24\pi\epsilon}(c-24)\nonumber\\
g_{\mu\nu} &\rightarrow &g_{\mu\nu}\exp [{1\over 2\pi\epsilon N}]\nonumber\\
V(N)&\rightarrow &V(N) +\Delta V(N)\nonumber\\
X^i &\rightarrow &X^i
\eea
where
\be
\Delta V(N)={V'\over 2\pi\epsilon}[1+{1\over 12}(c-24)]
\ee
 From the above equation we see that the theory is renormalisable
in the usual way  when $\Delta V$ is proportional to $V$. This
is the case for Liouville potentials of the form
$V(N)=\mu\exp [\alpha N]$. More generally, if the potential
depends on a set of bare constants $\alpha _i$, the theory
is renormalisable whenever $V'$ can be written as a linear
combination of ${\partial V\over\partial\alpha _i}$.
In this situation, the divergence can be
absorbed into the bare constants.

For other  potentials, the theory
is renormalisable only in a generalised sense. For example, in the
Jackiw Teitelboim model the linear potential $V(N)=\Lambda N$
gets an $N$-independent renormalisation, and the infinities cannot
be absorbed into the bare constant $\Lambda$. However, if one considers
$V(N)=\Lambda N+\mu$, the divergence can be absorbed into the
constant $\mu$.

\setcounter{chapter}{6}
\setcounter{section}{1}
\setcounter{section}{1}

\subsection*{6.Conclusions}

In this paper we showed that the Jackiw-Teitelboim model coupled
to Liouville theory and $c$ scalar fields can be interpreted as a
theory of critical strings as well as a theory of non-critical strings.
The target space in which the critical string propagates has dimension
$D=2+c$.

We then considered a more general dilaton-gravity theory and analysed it
from a perturbative point of view.
Some of the results of the conformal field theory were
reproduced at the one loop level. In particular, the anomalous
dimensions of certain operators do not get renormalised owing to the
absence of any poles in the graviton propagator.
Furthermore, the coefficient of the anomaly term (the term
proportional to $R$) in the one loop divergence is exactly
equal to the coefficient appearing in front of the Liouville
action of the non-critical string.

We have also analysed the renormalisability of the dilaton-gravity
theories. We showed that the
Liouville potentials are privileged because they produce theories
which are on shell renormalisable, i.e., the infinities can
be absorbed into the bare constants
of the theory and field redefinitions. For other potentials,
the renormalisation of the
theory implies a change in the functional form of the potential,
as the tachyon renormalisation in the non-linear sigma model.
These results agree with previous ones obtained in
the conformal gauge [27].

Let us now comment on previous calculations in background gauges
done by us and other authors [21,28,29]. The results we
obtained here correct our  claims in an earlier version of
this work  about the
$\beta$-dependence of the one loop divergence.
The one loop
counterterm has also been computed in ref. [28]. Unfortunately,
the results obtained there are different and seem not to
imply renormalisability for Liouville potentials.  A possible source of
disagreement may be  the fact that the doubling trick is lacking in that
calculation. The same problem appears in refs. [21,29], where
the one loop divergences were computed for the non-local
Polyakov action.

Finally, we mention that interesting developments would be
to compute the Vilkovisky DeWitt effective action [30],
as well as the analysis of the supersymmetric version of
the Jackiw Teitelboim model.
\vspace {2cm}
\par
{\bf Acknowledgements:}
We would like to thank
C. Marzban, S. Odintsov, J. G. Russo and specially
A.A. Tseytlin for helpful
discussions and correspondence.
We would like also to thank IAEA and UNESCO for
financial support.
\newpage

\setcounter{chapter}{7}
\setcounter{section}{1}
\setcounter{section}{1}

\subsection*{7. Appendix}

\setcounter{equation}{0}
In this appendix we will prove the formula (4.2). Consider the
Green function $G(m^2,x,x')$ defined by
\be
[-\nabla^2 + A^{\mu}\nabla_{\mu} +  M + m^2 ]
G(m^2,x,x')=\delta ^d(x,x')
\ee
where $m$ is a mass which will dissappear at the end of the
calculation. We want to compute the divergent part of
\be
ln \ det [ -\nabla^2 + A^{\mu}\nabla_{\mu} +  M + m^2] =-ln \ det G=
-Tr\ ln G
\ee
in the limit $d\rightarrow 2$.

Inserting the proper time representation
\be
G(m^2,x,x')=\int_0^{\infty}ds K(x,x',s)
\ee
into Eq.(7.1) we find the following Schroedinger equation for
the kernel $K(x,x',s)$
\be
\partial_s K=-[-\nabla^2 + A^{\mu}
\nabla_{\mu}+M+m^2]K
\ee
with the boundary condition $K(x,x',0)=\delta^d (x,x')$.

According to the  standard Schwinger DeWitt technique we write [23]

\be
K(x,x',s)={1\over (4\pi s)^{d/2}}\exp-[{\sigma (x,x')\over 2s}+m^2s]
\Delta ^{1/2}(x,x')\Omega(x,x')
\ee
where $\sigma (x,x')$ is one half the squared geodesic distance
between $x$ and $x'$ and
\newline
$\Delta (x,x')g^{1/2}(x)g^{1/2}(x')$
is the Van Vleck-Morette determinant. The Schroedinger equation
for the kernel then becomes
\be
\partial_s\Omega +{\sigma^{;\mu}\Omega_{;\mu}\over s}-
{A_{\mu}\sigma^{;\mu}\over 2s}\Omega+\Delta^{-1/2}[-\nabla^2+
A_{\mu}\nabla^{\mu}+M](\Delta^{1/2}\Omega )=0
\ee

Expanding $\Omega (x,x',s)$ in powers of $s$
\be
\Omega (x,x',s)=\sum _{k\ge 0}s^ka_k(x,x')\,\,\, ,
\ee
we get a set of recursion relations for the functions $a_k(x,x')$.
We will need only the first two which are
\bea
0&=&\sigma^{;\mu}a_{0;\mu}-{1\over 2}a_0 A_{\mu}\sigma^{;\mu}\\
0&=&a_1 + \sigma^{;\mu}a_{1;\mu} - {1\over 2}a_1 A_{\mu}\sigma^{;\mu}
\nonumber\\
&+& \Delta^{-1/2}[-\nabla^2+
A_{\mu}\nabla^{\mu}+M](\Delta^{1/2}a_0)
\eea
with the boundary condition $[a_0]=1$ (the brackets denote
the coincidence limit).

Using the representation
\be
ln\ det G=Tr\ ln G =\int_0^{\infty}{ds\over s}  Tr K
\ee
we find that the divergent part is independent of $m$ and given by
\be
ln\ det G|_{div}={1\over 2\pi\epsilon}
\int d^dx \sqrt g[tr a_1]
\ee
It is then necessary to extract $[a_1]$ from the equations (7.8) and (7.9).
As usual, this can be done taking derivatives and coincidence
limits of these equations. The following identities are
useful [31]
\bea
&[\sigma]=0\,\,\,\, [\sigma_{\alpha_1...\alpha_{2k+1}}]=0
\,\,\,\, [\sigma_{\alpha\beta}]=g_{\alpha\beta}&\nonumber\\
&[\Delta^{1/2}]=1\,\,\,\, [\Delta^{1/2}_{;\alpha}]=0\,\,\,\,
[\Delta^{1/2\,\,\, ;\alpha}_{;\alpha}]={1\over 6}R&
\eea

The coincidence limit of Eq.(7.9)  gives
\be
[a_1]=-M-A_{\mu}[a_0^{;\mu}]+{1\over 6}R+[\nabla^2 a_0]
\ee
Taking the covariant derivative of Eq.(7.8) and then the
coincidence limit we find
\be
[a_0^{;\mu}]={1\over 2}A^{\mu}
\ee
Doing the same with the second covariant derivative we obtain
\be
[\nabla^2a_0]={1\over 4}A_{\mu}A^{\mu}+{1\over 2}\nabla_{\mu}
A^{\mu}
\ee
Combining Eqs.(7.13) to (7.15) we get, up to total derivatives,
\be
[a_1]={1\over 6} R-{1\over 4}A_{\mu}A^{\mu}-M
\ee

Replacing Eq.(7.16) into Eqs.(7.11) and (7.2) we obtain the desired result
Eq.(4.2).


\newpage




\begin{thebibliography}{99}
\bibitem{any}
A. M. Polyakov, Mod. Phys. Lett. A2 (1987) 893.
\bibitem{any}
V. G. Knizhnik, A. M. Polyakov and A. B. Zamolodchikov, Mod. Phys. Lett. A3
(1988) 819.
\bibitem{any}
J. Distler and H. Kawai, Nucl. Phys. B321 (1989) 509.
\bibitem{any}
F. David, Mod. Phys. Lett. A3 (1988) 1651.
\bibitem{any}
V. Kazakov, Phys. Lett. B150 (1985) 282; Mod. Phys. Lett. A4 (1989) 2125;
\newline
A. Ambjorn, B. Durhuus and J. Fr\"ohlich, Nucl. Phys. B257 (1985) 433;
\newline
V. Kazakov, I. Kostov and A. A. Migdal, Phys. Lett. B157 (1985) 295;\newline
D. Boulatov, V. Kazakov, I. Kostov and A. A. Migdal, Nucl. Phys. B275 (1986)
641;\newline
J. Ambjorn, B. Durhuus, J. Fr\"ohlich and P. Orland,
Nucl. Phys. B270 (1986)
457.
\bibitem{any}
E. Br\'ezin and V. Kazakov, Phys. Lett. B236 (1990) 144.
\bibitem{any}
M. Douglas and S. Shenker, Nucl. Phys. B325 (1990) 635.
\bibitem{any}
D. J. Gross and A. A. Migdal, Phys. Rev. Lett. 64 (1990) 127;
Nucl. Phys. B340
(1990) 333.
\bibitem{any}
H. Kawai and M. Ninomiya, Nucl. Phys. B336 (1990) 115.
\bibitem{any}
I. Jack and D. R. T. Jones, Nucl. Phys B358 (1991) 695.
\bibitem{any}
R. Jackiw, in Quantum Theory of Gravity, ed. S. Christensen (Hilger, 1984);
\newline
C. Teitelboim, in Quantum Theory of Gravity, ed. S. Christensen (Hilger, 1984);
Phys. Lett. B126 (1983) 41.
\bibitem{any}
A. H. Chamseddine, Phys. Lett. B256 (1991) 379.
\bibitem{any}
F. D. Mazzitelli and N. Mohammedi ,  Phys. Lett. B268 (1991) 12.
\bibitem{any}
M. Henneaux, Phys. Rev. Lett. 54 (1985) 959.
\bibitem{any}
J.D. Brown, M. Henneaux and C. Teitelboim, Phys. Rev. D33 (1986) 319.
\bibitem{any}
T. Fukuyama and K. Kamimura, Phys. Lett. B160 (1985) 259.
\bibitem{any}
A. H. Chamseddine and D. Wyler, Phys. Lett. B228 (1989); Nucl. Phys. B340
(1990) 595;\newline
K. Isler and C. A. Trugenberger, Phys. Rev. Lett. 63 (1989) 834.
\bibitem{any}
N. Mohammedi, Mod. Phys. Lett. A5 (1990) 1251.
\bibitem{any}
A. A. Tseytlin, "On the tachyonic terms in the string effective action",
preprint JHU-TIPAC-91004.
\bibitem{any}
A. H. Chamseddine, "A study of non critical strings in arbitrary dimensions",
preprint ZH-TH-13/1991
\bibitem{any}
S. Ichinose, Phys. Lett. B251 (1990) 49.
\bibitem{any}
T. Banks and O'Laughlin, Nucl. Phys. B362 (1991), 649;
\newline
C. Callan, S. Giddings, J. Harvey  and A. Strominger, Phys. Rev. D45,
R1005 (1992)
\newline
T. Banks, A. Dabholkar, M.  Douglas and M. O'Loughlin,
Phys. Rev D45, 3607 (1992);
\newline
S. Hawking, "Evaporation of two dimensional black holes",
preprint CALT 68, 1774 (February 1992);
\newline
J.G. Russo, L. Susskind and L. Thorlacius, "Black hole evaporation in
1+1 dimensions", preprint SU-ITP-92-4 (January 1992);
\newline
L. Susskind and L. Thorlacius, "Hawking radiation and back-reaction",
preprint SU-ITP-92-12 (March 1992).
\bibitem{any}
B. S. DeWitt, Dynamical Theory of Groups and Fields, (Gordon and Breach, 1965).
\bibitem{any}
B.S. DeWitt, Phys. Rev. 162 (1967) 1195.
\bibitem{any}
G. 't Hooft and M. Veltman, Ann. Inst. H. Poincar\'e 20,1 (1974) 69.
\bibitem{any}
G. 't Hooft, in Recent Developments in Gravitation, Cargese 1978,
eds. M. Levy and
S. Deser (Plenum Press, New York and London 1979).
\bibitem{any}
J. G. Russo and A.A. Tseytlin, "Scalar-tensor quantum gravity in two
dimensions",
preprint SU-ITP-92-2, DAMTP-1-1992.
\bibitem{any}
S. D. Odintsov and I. L. Shapiro,  Phys. Lett. B263 (1991) 183;
\newline
preprint FTUAM 91-33 (October 1991).
\bibitem{any}
S. Ichinose,
preprint YITP/K-876 (August 1990).
\bibitem{any}
R. Kantowski and C. Marzban, in preparation
\bibitem{any}
A. O. Barvinsky and G. A. Vilkovisky, Phys. Rep. 119 (1985) 1.

\end{thebibliography}
\end{document}